\documentclass[aps,prl,twocolumn,nofootinbib,showpacs,superscriptaddress,groupedaddress]{revtex4-1}
\usepackage{amsmath,amssymb,graphicx,mathrsfs,slashed,caption}
\usepackage{lipsum, babel}
\usepackage{subcaption}
\usepackage[usenames, dvipsnames]{color}
\usepackage{tabularx}
\usepackage[
verbose,
colorlinks=true,
naturalnames=true,
linkcolor=blue,
citecolor=black,
]{hyperref}
\newcommand{\be}{\begin{equation}}
\newcommand{\ee}{\end{equation}}
\newcommand{\bea}{\begin{eqnarray}}
\newcommand{\eea}{\end{eqnarray}}

\def\({\left(} \def\){\right)}

\begin{document}
\title{{\Large 
Tidal-heating/viscous dissipation correspondence in black holes and viscous compact objects }}

\author{\large Yotam Sherf }
\affiliation{\ \\ Department of Physics, Ben-Gurion University,
	Beer-Sheva 84105, Israel \\ 	sherfyo@post.bgu.ac.il}

	\begin{abstract}

		\vspace{0.2cm} 
			The effect of energy absorption during the binary evolution of Exotic-compact-objects (ECOs) is extensively studied.  We review the underlying mechanism that provides the energy dissipation in material objects - tidal friction. We show that unlike typical astrophysical objects, where absorption due to viscosity is negligible, in ECOs, absorption could potentially  mimic the analogous effect of black-holes (BHs) - tidal heating. We stand for their differences and similarities in the context of energy dissipation during the inspiral. Inspired by the membrane paradigm and recent studies, we demonstrate how viscosity is a defining feature that quantifies how close is the ECO absorption to that of a classical BH absorption.  We show that for ECOs, viscosity can induce significant modifications to the GW waveform, which in some favorable scenarios of super-massive binaries of equal mass and spin, enables the measurement of the ECO absorption in the future precision gravitational-wave (GW) observations. Finally, we discuss the implications on the ECO reflection coefficient and the relation to the universal viscosity to volume entropy bound.

	\end{abstract}
		\maketitle
	\newpage
	\renewcommand{\baselinestretch}{1.0}\normalsize

	\begin{subequations}
		\renewcommand{\theequation}{\theparentequation.\arabic{equation}}
				\section{Introduction}\label{int}
Inspired by current observations from LIGO \cite{TheLIGOScientific:2016src,TheLIGOScientific:2017qsa,Abbott:2020uma} and future Laser Interferometer Space Antenna (LISA) \cite{LISA,AmaroSeoane:2012je}, the era of gravitational-wave (GW) spectroscopy
entails an outstanding opportunity for strong gravity tests with
precision GW measurement and to examine the consistency of Einstein's General-Relativity (GR) with GW observations.
One of the most tantalizing GR predictions is the existence of BHs and their event horizon, a surface characterized by the discontinuity of the causal structure and possess the unique property of being a perfect absorber. Thus, any deviations aside from the ultimate absorption support the possibility of the existence of ECOs and the emergence of new physics \cite{Cardoso:2016rao,Cardoso:2017cqb}.

In GR, the emergence of new physics
is expected to appear in the strong-field regime, when approaching the UV cutoff. At this regime, modifications to GR arise naturally, and additional quantum scales such as the string scale enters \cite{Boulware:1985wk,Zwiebach:1985uq}. Furthermore, the existence of these quantum motivated dark compact object is necessary  for  a  consistent  description of  evaporating  BHs as well as other open questions \cite{Braunstein:2009my,Brustein:2013ena}.
Among the quantum inspired ECOs alternatives are known BH polymers \cite{Brustein:2016msz,Brustein:2017koc}, fuzzballs \cite{Skenderis:2008qn}, Firewalls \cite{Almheiri:2012rt,Mathur:2009hf}, gravastars \cite{Mazur:2001fv} in addition to other quantum effects that were recently studied \cite{Giddings:2017mym,Cardoso:2019apo,Wang:2019rcf,Brustein:2019twi,Brustein:2020tpg,Agullo:2020hxe}, (see also \cite{Cardoso:2019rvt} for an extensive review).
Moreover, recent GW observations do not exclude the existence of dark compact objects \cite{LIGOScientific:2019fpa,Abbott:2020khf}.  Therefore, the corrections demonstrated by these beyond GR models, especially those that modify the BH absorption, should be taken into account when performing an ECO search. 

During the inspiral evolution of BHs, and in particular at its early stages, the GW emission is governed by the point mass orbital motion. At later stages, the mutual tidal interaction of the companions becomes significant, this gives rise to a dissipation of energy and angular momentum, a phenomenon known as tidal heating \cite{Hawking:1972hy,Hartle:1973zz,Thorne:1984mz,Poisson:1994yf,Poisson:2004cw,Alvi:2001mx}.
The effect of tidal heating in BHs can be illustrated from a Newtonian dynamic perspective \cite{Thorne:1986iy,Poisson:2009di,Cardoso:2012zn}. Where one of the companions is an ECO, characterized by an average kinematic viscosity $\nu_{_{ECO}}$, then dissipation of energy and angular momentum are attributed to tidal work and tidal torque exerted by the external tidal field on the body.
The tidal heating - Newtonian dynamics correspondence is best emphasized from a thermodynamic picture \cite{Poisson:2009di}. Where for a fluid star, energy is dissipated into heat according to $\dot{Q}=\dot{W}-\Omega_{_{ECO}} \dot{J}$, which means that heat generation is attributed to the tidal work ($\dot{W}$) and to the angular momentum flow $(\Omega_{_{ECO}} \dot{J})$ where $\Omega_{_{ECO}}$ is the body's angular velocity. In the analogous BH thermodynamic description we have $(\kappa/8 \pi)\dot{A}=\dot{M}-\Omega_{_{BH}} \dot{J}$, where $T_{BH}=(\kappa/2\pi)$ is the BH  temperature, $\kappa$ is its surface gravity and $S_{BH}=A/4$ is the BH entropy. Then, from the second law of thermodynamic $\delta Q = T \delta S$, we find that BHs dissipates energy in similar to a Newtonian viscous body.

The uniqueness of the effect is revealed when observing the frequency content of the heating formula. From \cite{Alvi:2001mx,Poisson:2004cw} tidal heating takes $\dot{M}\propto  \Omega(\Omega-\Omega_{_{BH}})$ and the rotational energy flow  $\Omega_{_{BH}} \dot{J}\propto \Omega_{_{BH}}(\Omega-\Omega_{_{BH}})$, then dissipation into heat reads $\dot{Q}\propto (\Omega-\Omega_{_{BH}})^2$, where $\Omega$ is the orbital angular velocity. At low spins, $\Omega_{_{BH}} \ll \Omega$ heat generation is governed by $\Omega$ and gets smaller as the body's spin increase where exact cancellation occurs  for  $\Omega_{_{BH}}=\Omega$, this threshold case is known as tidal locking. 
At higher spins, $\Omega_{_{BH}}\gg \Omega$, which is a typical situation at the inspiral phase, $\dot{Q}\propto \Omega_{_{BH}}^2$, and the entire effect is dominated by the object's spin. Moreover, since $\dot{J}<0$ the body's rotational energy flows outward and angular momentum conservation implies that spin angular momentum is removed to the orbit. The phenomena of energy extraction from the rotating body is the underlying mechanism for superradiance \cite{Cardoso:2012zn,Brito:2015oca}.

Here we study the viscous dissipation at the inspiral phase, which as previously mentioned, since the inspiral time scale is longer than the binary time scale $\Omega^{-1}\ll 1/R \sim \Omega_{_{ECO}}^{-1}$, viscous dissipation is governed by $\Omega_{_{ECO}}$. Motivated by the membrane paradigm \cite{Thorne:1986iy} and  the similarity shared by the tidal interaction of BHs and Newtonian viscous bodies. We show that in some cases dissipation in ECOs is comparable to that of a BHs and due to the large dephasing can produce an observable imprint
on the GW waveform as demonstrated in different context at \cite{Datta:2019euh,Datta:2019epe,Datta:2020rvo,Maselli:2017cmm}, which eventually could be detected with LISA. 
\section{Tidal heating - viscous dissipation correspondence}\label{Fra}
In the current setup we consider the inspiral phase of a binary system, whose companions are ECOs of comparable masses $M$ and radii $R$. The objects revolve on a circular orbit of radius $b\gg M$. At this phase the companions are tidally interacting, the mutual interaction is characterized to leading order by the quadrupolar tidal field $\mathcal{E}_{ab}$, which for Newtonian theory is defined in terms of the external Newtonian potential by $\mathcal{E}_{ab}=\partial_a \partial_b U_{ext}$. As a result of the slowly varying induced tidal field, the material body develops, to leading order, quadrupole  moment, given asymptotically in Cartesian coordinates by $Q_{ab}=\int \rho d^3x (x_ix_j-\frac{1}{3}r^2\delta_{ij})$, in which $\rho$ is the mass density. The strength of the response to the external tidal field depends on the body composition and is quantified in terms of the tidal Love number, which relates the induced quadrupole moment to the quadrupolar tidal field by $Q_{ab}=-\frac{2}{3}k_2R^5\mathcal{E}_{ab}$. Here $k_2$ is the dimensionless $l=2$ electric tidal Love number \cite{Hinderer:2007mb,Damour:2009vw,Binnington:2009bb} and $R$ is the stars radius.

 In the absence of viscosity, the tidal forces produced by each of the companions raise a bulge in the object,  the height of the bulge and its position are determined by the external tidal field. The relation of the bulge to the external tidal field is obtained by demanding that the potential energy gained by the raised bulge $\Delta R$, is conserved with respect to the external potential such that $ (M/R)\Delta R\sim U_{ext}$, where $U_{ext}$ is evaluated at the tides position. Then, expanding the Newtonian potential in $b \gg R$ we get $ U_{ext}\sim \mathcal{E}R^2$, where $\mathcal{E}= M_{ext}/b^3$ is the magnitude of the tidal field. Eventually, the height of the bulge takes the form $\Delta R (t)\sim\mathcal{E}(t)R^3$, where the time dependence indicates that the object response to the applied tidal field is instantaneous\footnote{	For a more detailed analysis regarding the relation of $\Delta R$  to $\mathcal{E}$ see  at \cite{Hartle:1974gy,Damour:2009va} in the context of the shape Love number}. 
 However, when the body's internal properties are considered, the effect of viscosity might be non-negligible. In this case, the response of the body to the exerted tidal force is not instantaneous and it suffers a time lag, characterized by the viscous time delay $\tau$. As a result, the tidal bulge is misaligned with the companion position, an effect that  displays a time delay in the induced quadrupole moment \cite{Poisson:2009di}, which for slowly-spinning becomes
\begin{align}
Q_{ij}(t)&=-\frac{2}{3}k_2R^5 \mathcal{E}_{ij}(t-\tau)\label{1.2}~,	
\end{align}
where $\tau$ is the viscous time delay. For the case of our interest we assume that the time scale associated with changes in the tidal field, given by $\Omega^{-1}\sim \mathcal{E}^{-1/2}$, where  $\Omega^{-1}$ is the orbital evolution time scale, is long  compared to the viscous time delay $\tau \ll \Omega^{-1}$. Then, the tidal field can be expanded into
\begin{align}
Q_{ij}(t)\approx -\frac{2}{3}k_2R^5 \left(\mathcal{E}_{ij}-\tau\dot{\mathcal{E}}_{ij}\right) .\label{1.21}	
\end{align}
 In general, the dependence of the viscous time delay in the internal properties is given by \cite{Poisson:2009di}
\begin{gather}
\tau~=~\alpha\nu~. \label{td}
\end{gather}
Where $\nu$ is the average kinematic viscosity and $\alpha$ is a dimensionless coefficient that depends on the objects interior. For example, an isotropic incompressible viscous body with compactness $C$ has $\alpha=\alpha_0=\frac{19}{2 C}$ \cite{Poisson:2009di}. For anisotropic fluids $\alpha=\alpha_0/\mathcal{A}$, where $\mathcal{A}$ is the anisotropy factor, with $\mathcal{A}=1$ denoting isotropic fluid \cite{Glampedakis:2013jya}. The importance in considering ECOs composed of  anisotropic matter is due to their natural appearance at high densities, in addition to their ability to sustain gravity above the Buchdahl bound $C>4/9$ \cite{Raposo:2018rjn}, and up to the BH limit $C=1/2$ \cite{Glampedakis:2013jya}. 

For the BH viscous delay,  we follow the membrane paradigm \cite{Thorne:1986iy} where the BH horizon is treated  as  one way viscous membrane with an effective kinematic viscosity of $\nu_{_{BH}}=M_{_{BH}}$. We find that for BHs, where the only relevant scale is the BH scale  $\tau_{_{BH}}=R_{_{BH}}$, so $\alpha=2$. 
The result is also consistent with the calculation of the phase lag of the bulge relative to the external companion position \cite{Hartle:1974gy}, where for non-spinning BHs $\phi= 4 M \Omega$, then since $\phi= \omega
 \tau$, and $\omega=2\Omega$ is twice the orbital frequency, we obtain $\tau_{_{BH}}=R_{_{BH}}$. Besides, it is important to mention that 
 although, as pointed out in \cite{Raposo:2018xkf}, horizon scale corrections leave an indistinguishable imprint on the spacetime of ECOs in comparison to that of BHs, the behaviour of the BH bulge is different than the ECO bulge.
 The difference in the bulge behaviour is not related to the compactness of the objects, nor how similar their spacetimes are, rather, it is related to the definition of the event horizon - a globally defined surface, where no light rays can escape from it to spatial infinity. Thus,  no local physical experiment can locate the event horizon \cite{Ashtekar:2004cn}. Owing to this property, the boundary conditions imposed on the event horizon are teleological \cite{Thorne:1986iy}. As a result, the bulge displays an a-causal behaviour, which means that in contrast to material bodies, it leads the orbit when $\Omega>\Omega_{_{BH}}$ and lags when $\Omega<\Omega_{_{BH}}$. This a-causal behaviour guarantees the stability of the hole under perturbations, as shown in  \cite{Thorne:1986iy}, where causal evolution leads to an exponential expansion of the BH and eventually destroys it (see also \cite{Hawking:1971vc} for a similar discussion in the context of BHs area theorem). 
 Moreover, the ECO considered here are horizonless, and so their outer surface is locally defined and smoothly connected to time-like geodesics.


Here we consider ECOs whose outer surface positioned around their photonsphere $R\sim 3M$ \footnote{In the next section we extend the results for Ultra-compact-objects $R=2M(1+\epsilon)$, which requires full GR treatment.}. 
In the presence of viscosity, the work exerted on the star by the external tidal field is partially converted into heat (tidal friction) and partially leads to  angular momentum flow (tidal torque exerted on the bulge). 
this is also reflected by the energy balance formula $\dot{E}_\nu=\dot{Q}+\Omega_{_{ECO}}\dot{J}$ (we  label $\dot{W}=\dot{E}_\nu$). Where in the case of slowly-spinning objects ($\Omega \gg \Omega_{_{ECO}}$), most of the work exerted by the tidal field
is dissipated into heat by the interior viscosity.
In Newtonian theory, the work applied by the tidal field on a fluid element $dm$ is $E_\nu=-\frac{1}{2}\int \mathcal{E}_{ab}x^ax^bdm$. For the rate of tidal work, differentiation and integration by parts yields 
\begin{gather}
\dot{E}_{\nu}~=~\dfrac{1}{2}Q_{ij}\dot{\mathcal{E}}^{ij}~.
\label{tdw}
\end{gather}
Then, substituting the induced quadrupole from Eq.~(\ref{1.21}) and dismissing total derivatives,  
 viscous dissipation reads \cite{Hartle:1974gy, Poisson:2009di}\footnote{The calculations are performed in $M_1$ local asymptotic rest frame (LARF1).}
\begin{gather}
\dot{E}_{\nu}~=~\dfrac{1}{3C^5}k_2\tau M^5\dot{\mathcal{E}}_{ij}\dot{\mathcal{E}}^{ij}~.
\label{eij}
\end{gather}
Where $C=M/R$ is the ECO compactness, $\dot{\mathcal{E}}=  \Omega M_{ext}/b^3$ and $\Omega = \sqrt{M_{ext}/b^3}$ is the orbital frequency. 

To proceed, we extend the relations of viscous bodies Eq.~(\ref{1.21})-(\ref{eij})  to GR BHs.
First, mentioning that for non-spinning BHs, the rate of tidal heating takes the  form of Eq.~(\ref{tdw}), $\dot{M}=\frac{1}{2}Q_{ij}^{_{BH}}\dot{\mathcal{E}}_{ij}$ \cite{Thorne:1986iy,Poisson:2004cw}, where the induced quadrupole moment of a mass $M$ BH is given by
\begin{gather}
	Q_{ij}^{_{BH}}~=~\dfrac{32}{45}M^6\dot{\mathcal{E}}_{ij}~,
	\label{qij}
\end{gather}
and is not proportional to the tidal field, as implied from the BHs no-hair properties. For the rate of tidal heating, substituting $Q_{ij}^{_{BH}}$ into $\dot{M}$ immediately reads (we label $\dot{M}=\dot{E}_{_{BH}}$)
\begin{gather}
\dot{E}_{_{BH}}~=~\dfrac{16}{45}M^6\dot{\mathcal{E}}_{ij}\dot{\mathcal{E}}^{ij}~.
\label{eth1}
\end{gather}  

The similarity of these effects is revealed when the viscous dissipation is given in terms of the tidal heating dissipation.
\begin{gather}
\dot{E}_{\nu}~=~\gamma \dot{E}_{_{BH}}~.
\label{eth}
\end{gather}
Where $\gamma$ is defined as the dimensionless effective absorption coefficient, that is given in terms of the viscous time delay $\tau$ Eq.~(\ref{td}) by
\begin{gather}
\gamma~=~ \dfrac{15\alpha}{32C^5} \dfrac{\nu}{M}k_2.
\label{g1}
\end{gather} 
The effective absorption coefficient $\gamma$ quantifies how close is the body absorption to that of a GR BH absorption, alternatively, how good is the body as a BH mimicker. By definition, $0\leq\gamma \leq1$, where $\gamma=0$ describes empty, nonviscous interior. On the contrary, $\gamma=1$ defines objects whose dissipation matches the BH dissipation. In general, for ordinary matter  $\gamma \ll 1$, to show this we consider cold neutron stars ($T\approx 10^6$K) that posses large viscosities $\nu{_{NS}}\approx 3 \times 10^{5} \frac{\text{m}^2}{\text{s}}$ \cite{LL1,LL2} (see Appendix for more details). To estimate  $\gamma$ from Eq.~(\ref{g1}), we take $\alpha_0$ and choose typical neutron star parameters $R=10^4$m$ , M=1.4M_{\odot}$ and tidal deformability parameter $k_2= 1/3$, the effective absorption for neutron stars reads
\begin{align}
	\gamma_{_{NS}}~&~\sim~ 9\times 10^{-5}  \times\left(\dfrac{\nu}{ 3\times 10^{5}\frac{\text{m}^2}{\text{s}}}\right)\left(\dfrac{1.4M_{\odot}}{M}\right)~.
	\label{g2}
\end{align}
Then, since $\gamma_{_{NS}}\ll1$ we conclude that for typical matter, viscous dissipation is negligible in comparison to the analogous effect in BHs. 
However, this might not be the case for exotic matter, in which the viscosity can be orders of magnitude larger than the value given in Eq.~(\ref{g2}). Eventually, in such scenarios, the resulted effective absorption becomes significant and approaches the BH absorption $\gamma \lesssim 1$.
 
 The motivation in considering matter providing non-negligible absorption, and hence large viscosity, is important when performing an ECO search, where any departure from the complete absorption indicate on the existence of ECOs. The existence of matter providing these properties  is inspired from both classical hydrodynamic and quantum mechanics perspective. Where large viscosity is attributed to a matter whose viscosity is of order of the analogous horizon viscosity \cite{Thorne:1986iy}, where according to the membrane paradigm, from $\nu_{_{BH}}=M$  
 \begin{equation}
 ~~~~~~\nu_{_{BH}}~\approx ~2.7\times 10^{13}\dfrac{m^2}{ s}\left(\dfrac{M}{65M_{\odot}}\right)~.
 \label{ebh}
 \end{equation}
  An example for such matter can be found in non-rotating strongly magnetized $(B=10^{15}\text{G})$ neutron stars, in which their viscous damping time $\tau_\nu \sim R^2/\nu$ is compared to the Alfven time scale, then, for a typical neutron star parameters we find $\nu \sim  4\times 10^{10} \text{m$^2$/s}$ \cite{Shapiro:2000zh,Yunes:2016jcc}. For boson stars \footnote{In most cases bosonic matter is assumed to have low viscosity.}, with radius $R\sim 3M$ and mass $65 M_{\odot}$, the damping time of the $l=2$ mode is $\tau\sim 320$ms, which yields a viscosity of $\nu\sim 2 \times10^{10}$m$^2$/s \cite{Yunes:2016jcc}. As for the quantum mechanical perspective,  motivated by the information paradox, where unitarization 
 requires non-local effects, or exotic matter that provides strong quantum effects that is outside of the standard model \cite{Brustein:2013ena,Brustein:2016msz,Giddings:2017mym,Buoninfante:2019swn,Giddings:2011ks,Almheiri:2012rt}. These theories involves a new scale of order of the Planck scale (as in string theory), which leads to deviation from the horizon and thus, display small corrections to the would be BH entropy $S_{_{ECO}}=A/(4g l_P^2)$, where $g \gtrsim 1$  is a new dimensionless scale of order unity or more, as predicted in string theory $g \sim l_s^2/l_P^2$, with $l_s$ being the string scale \cite{Dienes:1996du}. The link to viscosity is made by the requirement that the exotic matter saturates the universal shear viscosity to entropy volume density  (see also Sec.~\ref{tif} for more details), $4 \pi{\eta}/{s}=1$. Then since $\eta= \rho \nu $, where $\rho$ is the mass density, we find that the ECO viscosity is suppressed in comparison to the BH viscosity $\nu_{_{ECO}}\sim \nu_{_{BH}}/g~$\cite{Thorne:1986iy,Brustein:2017koc,Kovtun:2004de}. The quantum perspective is discussed separately in Sec.~\ref{uco}.
 
 At this stage, motivated by the vast BH alternatives, we turn to the evaluation of the ECO effective absorption Eq.~(\ref{g1}). From the analogous BH viscosity $\nu_{_{BH}}=M$, we get  $\gamma_{_{ECO}}=\beta \nu/\nu_{BH}$. 
 Where $\beta $ depends on the ECO interior and can be found by specifying the ECO equation-of-state (EOS). For example, we notice that for the previous exotic alternatives suggested $\nu \sim 10^{-3}\nu_{_{BH}}$, and extremely compact models with $C=1/3$ have $k_2\sim 10^{-2}$ and $\alpha \sim 10$ (see examples for several NS models at \cite{Hinderer:2007mb}). This leads to $\beta \nu \lesssim \nu_{_{BH}}$, which implies that viscous dissipation in ECOs is comparable to horizon dissipation in an equal mass BHs. 
 Here we do not examine the specific details of the ECO interior neither the underlying mechanism that provides large viscosity, rather, we are satisfied with knowing that such scenarios are physically viable. Hence we define the ECO effective viscosity $\beta \nu=\nu_{_{ECO}} $, and focus on scenarios where $\nu_{_{ECO}}/\nu_{_{BH}} \lesssim 1$ in addition to their implications on future GW observations.
 Then, we define the ECO effective absorption coefficient 
 \begin{gather}
\gamma_{_{ECO}}~=~  \dfrac{\nu_{_{ECO}}}{\nu_{_{BH}}}
\label{ge}~.
  \end{gather}
  Where $0\leq\gamma_{_{ECO}} \leq1$, and is maximized when $\nu_{_{ECO}}$ approaches $ \nu_{_{BH}}$, so $\gamma_{_{ECO}} \rightarrow\gamma_{_{BH}}=1$.
  
   The conclusion is that in general, viscous dissipation in ECOs can mimic the effect of tidal heating in BHs, and as we show in details below, under some favourable conditions can leave a significant imprints on the emitted GW waveform during the inspiral phase as well at the ringdown phase in the form of echoes when the ECO are very compact $R=2M(1+\epsilon)$, where $\epsilon\ll 1$.
 
 \textit{The effect of spin }
 In the previous section, we focused on the absorption of non-spinning ECOs. However, the inclusion of spin effects increases the energy dissipation at 1.5 PN relative order and therefore is essential to the detectability estimation of $\gamma_{_{ECO}}$. 
  
 The generalization of $\gamma_{_{ECO}}$ to the spinning case follows the procedure given above, where now $\dot{E}_{_{ECO}} ~\text{and}~ \dot{E}_{TH}$ are amplified due to the effect of rotation. The corresponding expressions for dissipation of energy in the spinning case can be found at \cite{Poisson:2004cw,Alvi:2001mx}. Here for simplicity we consider the case of a rigid rotation on equatorial orbits, then, to leading order, we find that Eq.~(\ref{tdw})  is valid and the relation given in Eqs.~(\ref{eth}) holds. Hence, to leading order in the spin, tidal dissipation is given by
 \begin{gather}
 \dot{E}_{i} ~=~ \gamma_{i}	\dfrac{32}{45}M^6\mathcal{E}^2(1+\sqrt{1-\chi^2})(1+3\chi^2)\Omega(\Omega-\Omega_{_{i}})~. \label{es1} 
 \end{gather}
 Where the subscript $i=(\text{BH,~ECO})$, $\Omega_{i}$ is the BH/ECO angular velocity,  $\gamma_{_{BH}}=1$. In addition, from the viscous dissipation of rotating stars given in \cite{Poisson:2009di} we obtain the spin-dependent ECO effective absorption 
 \begin{gather}
 	\gamma_{_{ECO}}(\chi)~=~\dfrac{\nu_{_{ECO}}}{\nu_{_{BH}}}\dfrac{2}{(1+\sqrt{1-\chi^2})(1+3\chi^2)}~.
 \end{gather}
 We first note that at low spins we recover the non-rotating results Eqs.~(\ref{eij}),(\ref{eth1}), and, from \cite{Berti:2004ny} we also know that at low spins the ECO metric can be approximated to that of a Kerr BH. Thus, the quadratic spin modifications to $\gamma_{_{ECO}}$ can be regarded as a relativistic corrections of the fluid velocity. For a viscous body, the velocity of a fluid element positioned at the outer surface $v_{_{R_+}}=R_+\Omega_{_{ECO}}\sim \chi $ (where $\Omega_{i}=\chi/2R_+$), so spin effects brings to $\dot{E}_{_{ECO}}$  relativistic corrections of order $(v/c)^2$
  \cite{Poisson:2009di}. Eventually, we conclude that the net effect of spin corrections is to reduce the non-rotating ECO effective absorption at high spins ($\chi\gtrsim 0.85$) by not more than a factor two, $\gamma_{_{ECO}}(\chi\gtrsim0.85) \sim 2 \gamma_{_{ECO}}$.


 
 	\end{subequations}\begin{subequations}
 		\renewcommand{\theequation}{\theparentequation.\arabic{equation}}

\section{Viscous dissipation in Ultra-Compact-Objects}\label{uco}
	In this section, we would like to extend the previous formalism to another sub-family of ECOs - Ultra-compact-objects (UCOs), in which their outer surface positioned at $R=2M(1+\epsilon)$ where $\epsilon\ll1$ (see classification at \cite{Cardoso:2019rvt}). 
Inspired by various quantum motivated models (see introduction and discussion below Eq.~(\ref{ebh})) and especially by the BH membrane paradigm, where the horizon is modeled as a one-way viscous membrane with shear viscosity $\eta_{_{BH}}=1/(16\pi)$. We extend the viscous description of BHs and generalize it to UCOs by considering horizon scale corrections and non-negligible viscosity $\eta_{_{UCO}}\lesssim \eta_{_{BH}}$.

The key point is the realization that the viscous dissipation of inspiraling UCOs is similar to the tidal dissipation of BHs \cite{Datta:2019epe},

 Eqs.~(\ref{eth1}),(\ref{es1}), since as pointed out at \cite{Raposo:2018xkf} the external geometry of UCOs can be well approximated by the Schwarzschild geometry (or by the Kerr geometry in the spinning case) up to some higher-order corrections in the perturbative parameter $\epsilon$. Therefore, Eq.~(\ref{es1}) is valid in addition to the correspondence Eq.~(\ref{eth}). The only difference is encoded in the UCO effective absorption coefficient $\gamma_{_{UCO}}$, which in contrast to the above reviewed ECOs, whose outer surface lies at $R\sim 3M$ so relativistic correction are small, the surface of UCOs
 lies deep inside their photonsphere and requires the full GR treatment.
 To proceed, we express Eq.~(\ref{es1}) in terms of the UCO reflection coefficient $\mathcal{R}$ \cite{Datta:2019epe},  
\begin{gather}
\dot{E}_{_{UCO}}~=~\left(1-|\mathcal{R}|^2\right)\dot{E}_{TH}~,
\end{gather}
and yields the relation $|\mathcal{R}|^2=1-\gamma_{_{ECO}}$.
This implies that a perfect absorber with $\gamma_{_{UCO}}=1$ has $\mathcal{R}=0$ which corresponds to $\eta_{_{UCO}}=\eta_{_{BH}}$, alternatively a perfect reflector has $\mathcal{R}=1$ which corresponds to negligible absorption $ \eta_{_{UCO}}\ll \eta_{_{BH}}$.

In order to find the UCO viscosity dependent reflection coefficient, we follow the derivation given in \cite{Maggio:2020jml} and extend their results for the inspiral phase where $M\omega\ll1$. We consider the quadrupolar $l=2$ perturbations, sourced by the external tidal field. Gravitational perturbation of the even-parity sector in the exterior Schwarzschild geometry are given by a simple wave-like differential equation \cite{Regge:1957td,Zerilli:1970se}
\begin{gather}
	\dfrac{d^2 \psi(x)}{dx^2}+\left( \omega^2- V\right)\psi(x)=0
\end{gather}
where the even-parity potential is given by
\begin{gather}
		V(r)~=~f(r)\dfrac{24 r^2(r+M)+18M^2(2r+M)}{r^3(2r+3M)^2}
\end{gather}
here $x$ is the familiar tortoise coordinate $dx/dr=1/f(r)$ and $f(r)=1-2M/r$. The scalar field $\psi$ obeys purely outgoing boundary conditions at infinity and a superposition of an incoming and a reflected outgoing waves near the UCO surface 
\begin{gather}
		\psi(x)~\sim~ e^{ i \omega x} \hspace{2.2cm} x\rightarrow \infty~~~\label{bc2}\\
	\psi(x)~\sim~ \mathcal{R}e^{ i \omega x}+e^{-i \omega x}~~~~~x\rightarrow x(R)\label{bc1}
\end{gather}
To relate the properties of the membrane to the external geometry and in order to compensate the discontinuity of the extrinsic curvature due to the viscous membrane, we impose a junction condition on the extrinsic curvature\footnote{ We  assume that the extrinsic curvature of the interior is zero \cite{Thorne:1986iy}.} \cite{Thorne:1986iy,Maggio:2020jml}
 \begin{gather}
 	Kh_{ab}-K_{ab}~=~8\pi T_{ab}~,\label{kj}
 \end{gather}
where $K_{ab}$ is the extrinsic curvature about the  3-dimensional membrane parametrized by the intrinsic coordinates $(t,\theta,\phi)$, $K=h_{ab}K^{ab}$ is the trace with respect to the induced metric $h_{ab}$ of the 3-dimensional membrane.
The stress-energy-momentum tensor (SEM) on the surface 
\begin{gather}
	T_{ab}~=~\rho u_au_b+(p-{\Theta}\xi)\gamma_{ab}-2\eta_{_{UCO}}\sigma_{ab}
\end{gather}
where $\rho, p, u_a$ are the density, pressure and the velocity of a fluid element on the membrane. $\xi, \Theta$ are the bulk viscosity and the expansion, $\gamma_{ab}=h_{ab}+u_au_b$ is the projection metric and $\sigma_{ab}$ is the shear tensor.
As demonstrated in \cite{Maggio:2020jml} (for similar approach see \cite{Oshita:2019sat}) by imposing the junction condition Eq.~(\ref{kj}) on the UCO surface, we get a boundary condition that relates the UCO reflection coefficient to the membrane properties 
\begin{gather}
	\dfrac{\psi'(x)}{\psi}=- i \dfrac{\nu_{_{UCO}}}{\nu_{_{BH}}} \omega +G(R,\omega,\nu, \xi) ~~~~x\rightarrow x(R).
	\label{bc}
\end{gather}
Where $G(R,\omega,\nu,\xi)$ is a function that depends on the membrane properties and is given in the Appendix.
We solve Eq.~(\ref{bc}) for the reflection coefficient near the UCO surface by imposing the boundary condition Eq.~(\ref{bc1}). Then, to first order in $\epsilon\ll1$ and for the low frequency tidal perturbation $M\omega\ll1$ that is related to the orbital frequency $\Omega=\sqrt{M/b^3}$ by $\omega=2\Omega$, the UCO reflection coefficient reads\footnote{The results agree with \cite{Maggio:2020jml} for $\epsilon\ll1$, where here we also neglected the bulk viscosity since it is insignificant to the absorption coefficient in the BH limit.}
\begin{gather}
|\mathcal{R}|^2\approx\left(\frac{1- \nu_{_{UCO}}/\nu_{_{BH}} }{1+\nu_{_{UCO}}/\nu_{_{BH}}}\right)^2\left(1-\frac{512 (16 \pi -3\nu_{_{UCO}}/\nu_{_{BH}} )  }{77 \left(1- \nu_{_{UCO}}^2/\nu_{_{BH}}^2\right)}\epsilon\right)~,
\label{ruc}
\end{gather}
Obviously, for $\nu_{_{ECO}}\ll1$ absorption is negligible and the UCO reflection is maximized $\mathcal{R}\approx1$. On the other hand, for $\nu_{_{UCO}}\approx \nu_{_{BH}}$ reflection is negligible $\mathcal{R}\ll1$. 

From the relation  $|\mathcal{R}|^2=1-\gamma_{_{ECO}}$ and to leading order, we obtain the UCO effective absorption coefficient 
\begin{gather}
	\gamma_{_{UCO}}~=~\dfrac{4 \nu_{_{UCO}}/\nu_{_{BH}}}{(1+\nu_{_{UCO}}/\nu_{_{BH}})^2}~.\label{guc}
\end{gather}
As expected, in the absence of viscosity $\gamma_{_{ECO}}=0$, alternatively, when $\nu_{_{UCO}}=\nu_{_{BH}}$ we recover the BH absorption $\gamma_{_{UCO}}=1$.
Furthermore, we notice that the effective absorption of ECOs Eq.~(\ref{ge}) is recovered in the BH limit, where for $\nu_{_{UCO}}/\nu_{_{BH}}\approx 1$ we find $\gamma_{_{UCO}}\approx \nu_{_{UCO}}/\nu_{_{BH}}$.

Finally, as previously explained, the UCO viscous dissipation is obtained by substituting $\gamma_{_{UCO}}$ into Eq.~(\ref{es1}).
Moreover, we argue that although the reflection coefficient is calculated for non-spinning UCOs, Eq.~(\ref{guc}) is still a good approximation for rotating UCOs. This claim is justified in the UCO corotating frame with angular velocity $\Omega_{_{UCO}}=\chi/(2M(1+\sqrt{1-\chi^2}))$, where the frequency of the external tidal perturbation is shifted to $\Omega-\Omega_{_{UCO}}$ such that at high spins $\chi\lesssim 1$  the low-frequency approximation breaks.
However, from Eq.~(\ref{bc}) we notice that higher orders in the frequency are coupled to $\epsilon$ through the function $G(\omega)$. Thus, since we are particularly interested in quantum corrections to the horizon where $\epsilon\ll1$, which is parametrically orders of magnitude smaller than the dimensionless spin parameter $\epsilon\ll\chi$. Higher-order spin corrections are always suppressed by $\epsilon$ and are unimportant to the absorption properties of the membrane.

\section{ Detectability}\label{det}
\begin{figure*}[t!]
	\centering
	\includegraphics[width=.492\linewidth]{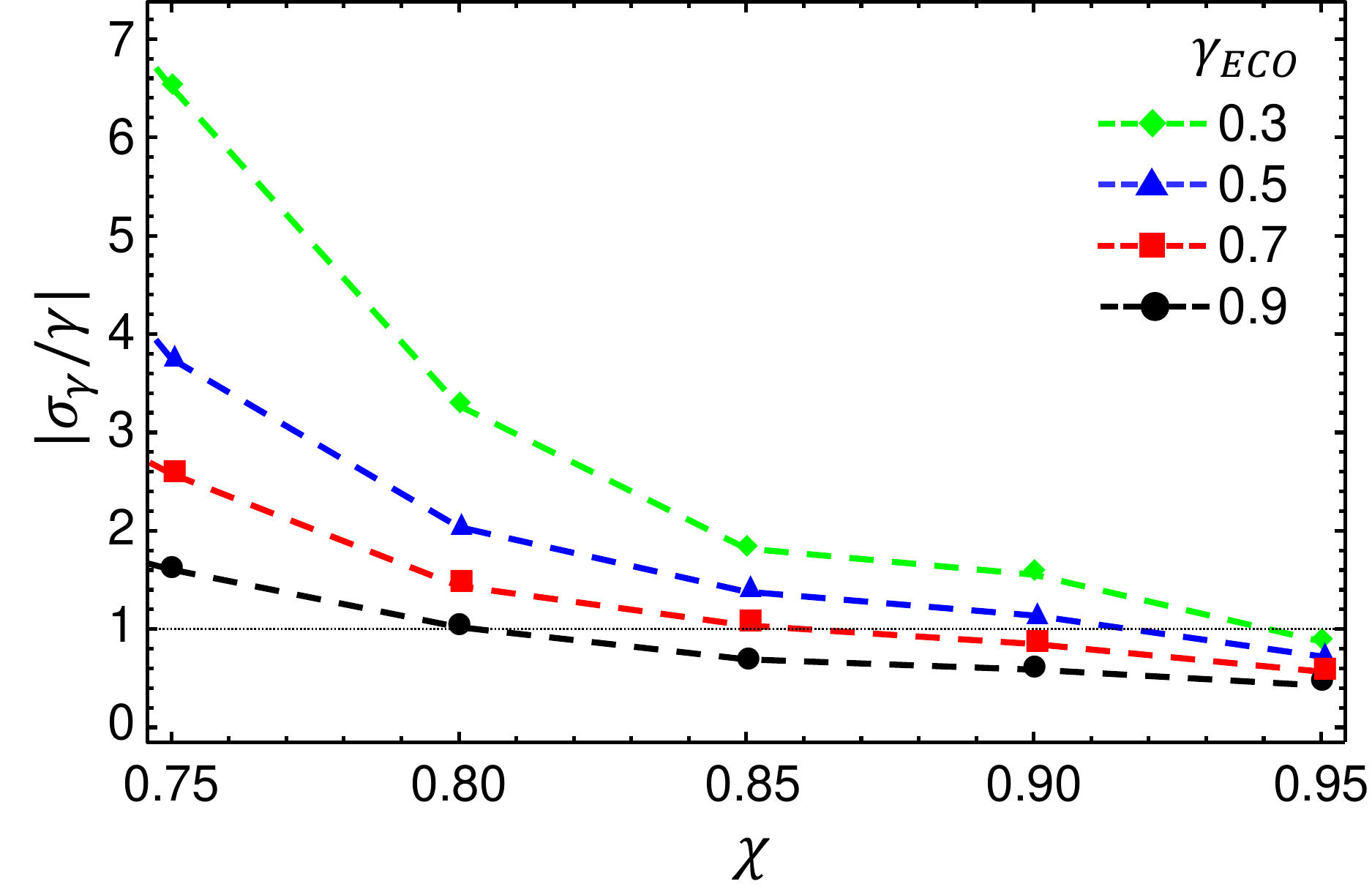}
	\includegraphics[width=.50\linewidth]{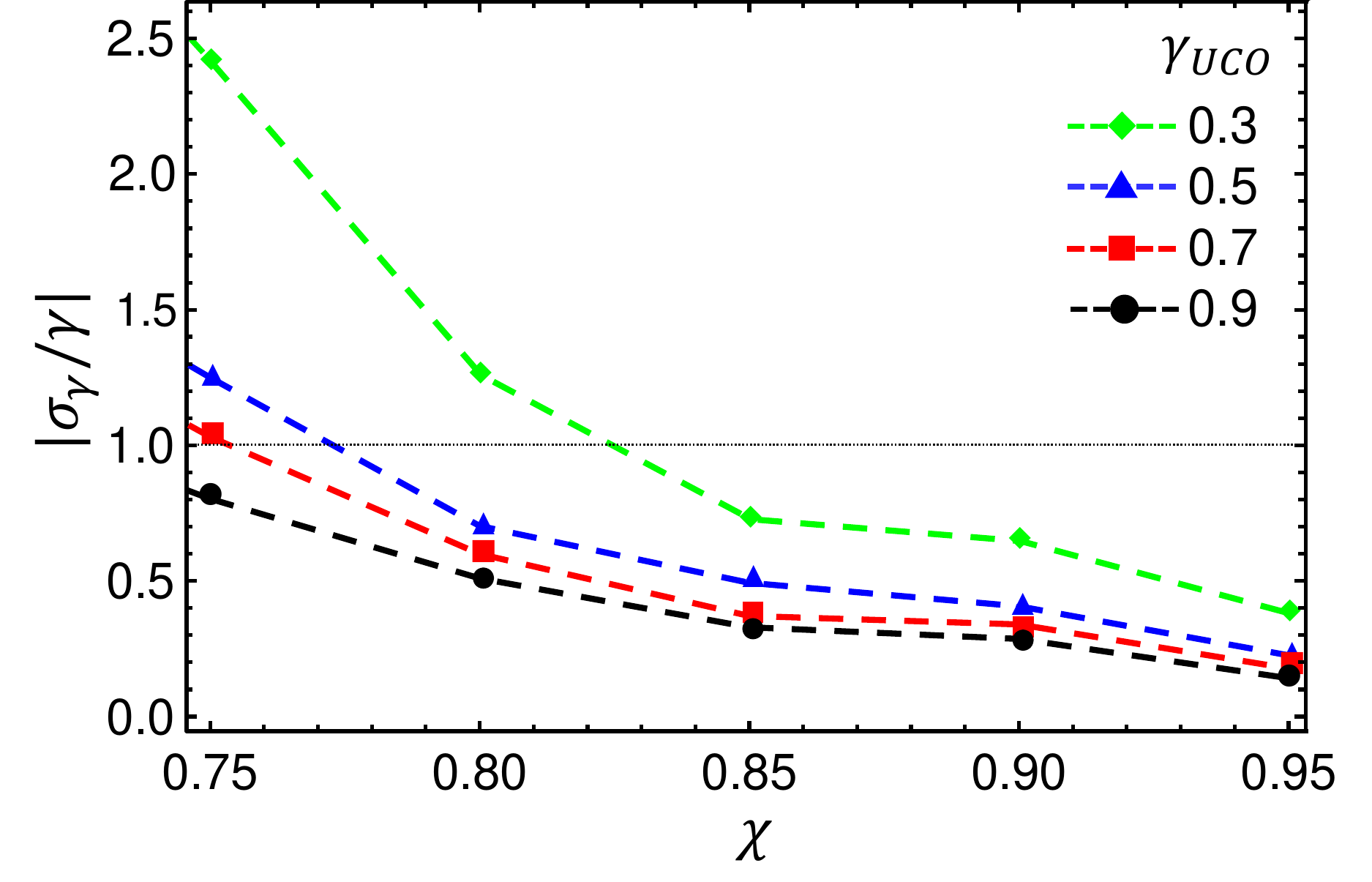} 
	\caption{{\small The relative statistical error vs  spin parameter for several values of $\gamma_{_{ECO/UCO}}$ Eqs.~(\ref{ge}),(\ref{guc}). We assume a binary of equal spin and nearly equal masses $M_1/M_2=1.01$ with luminosity distance $D_l=2$Gpc. Points below the horizontal dashed line correspond to detection at better than $1\sigma$ confidence. The $\gamma_{_{UCO}}=\left(0.3, 0.5, 0.7, 0.9\right)$ corresponds to $\nu_{_{UCO}}/\nu_{_{BH}}=\left(0.08,0.17,0.29,0.51\right)$.}}
	\label{s2}
\end{figure*}
In this section, we discuss the possibility of measuring the ECO effective absorption $\gamma_{_{ECO}}$ in future LISA observations of supermassive, highly-spinning equal-mass binaries. Which according to \cite{LISA}, LISA could track for a time period $T\sim 1$year and observe from the early stages of the inspiral and up to the coalescence with high SNR ($\gtrsim 10^3$). We show that for such binary systems, the statistical error due to the detector noise enables to set constraints on $\gamma_{_{ECO}}$, which according to Eq.~(\ref{ge}) constraints the properties of the exotic matter through $\nu_{_{ECO}}$.
When spin effects are considered, we find that the higher the spin of the rotating body, the lower the relative statistical error. The improvement in accuracy at large spins is mainly since tidal heating effects of spinning BHs are leading to the non-spinning case at 1.5 PN relative order, where the spinning (non-spinning) enters at 2.5 PN (4PN) relative to the leading order GW phase term. Another important reason is due to the signal-to-noise-ratio (SNR) sensitivity at the late stages of the inspiral, where the orbital frequency is large SNR$^2\sim \int_{\min}^{\max}\frac{\tilde{h}(f)^2}{S_n(f)}df$. Where $\tilde{h}(f)$ is the Fourier transform GW waveform and $S_n(f)$ is the detector noise spectral density. We find this effect to be extremely important for UCOs whose outer surface lies deep inside their photonsphere. We denote $f_{\max}=f_{\text{ISCO}}$, and $f^{\text{Kerr}}_{_{\text{ISCO}}}$ is Kerr's ISCO which for certain spins ($\chi \gtrsim 0.7$) modifies $f_{\text{max}}$ substantially $f^{\text{Kerr}}_{_{\text{ISCO}}}/f^{\chi=0}_{_{\text{ISCO}}} \approx 3$ \cite{Favata:2010ic}.

The analysis given below follows the analysis given in \cite{Maselli:2017cmm}, there it is
shown that highly-spinning super massive BHs characterized by maximal absorption $\gamma_{BH}=1$ can be distinguished from ECOs with zero absorption $\gamma_{_{ECO}}=0$, which means that the statistical error due to the detector noise is small enough and enables the measurement of the absorption coefficients with high certainty. However, here we consider the possibility that dissipation effects in ECOs can mimic the effect of dissipation at the horizon, as described by $\dot{E}_{_{ECO}}=\gamma_{_{ECO}} \dot{E}_{_{BH}}$. Therefore, as displayed in Eq.~(\ref{ge}), ECOs could, in principle, produce comparable effects to the BH absorption, namely  $\gamma_{_{ECO}}\lesssim1$. The results are displayed in Fig.~\ref{s2}.

To proceed, our aim is to estimate the statistical error in the measurement of $\gamma_{_{ECO}}$. We apply a parameter estimation method based on the Fisher information matrix that is accurate for large SNR events and described by a Gaussian distribution. Where for a given detector signal $s(t)=h(t,\theta^i)+n$, where $n$ is the noise and $h(t,\theta^i)$ is the model signal that depends on the parameters $\theta^i$. The posterior probability distribution can be approximated by 
\begin{gather}
	p(\theta^i|s) \propto e^{-\frac{1}{2}\Delta \theta^i \Delta \theta^j \Gamma_{ij}}~,
\end{gather} 
Where $\Gamma_{ij}$ is the Fisher matrix defined by 
\begin{gather}
	\Gamma_{ij}=\left(\frac{\partial h}{ \partial \theta^i}\Big|\frac{\partial h}{ \partial \theta^j}\right)~,
\end{gather}
and the inner product $(\cdot|\cdot)$ is defined by
\begin{gather}
	(h_1|h_2)~=~4~\text{Re}\int _{f_{min}}^{f_{max}} \frac{\tilde{h}_1(f)\tilde{h}_2^*(f)}{S_n(f)}df~.\label{hhi}
\end{gather}
Here $S_n(f)$ denotes LISAs noise spectral density \cite{LISA,Cornish:2018dyw}.  For the mass range considered below $M=10^6 M_\odot$ the minimal frequency is the entrance frequency to LISA's observation band is denoted by, $f_{min}\approx 10^{-5}$Hz which corresponds to observation time of about one year \cite{Berti:2004bd}. The maximal frequency for UCOs is taken to be the frequency at the Kerr's ISCO $f_{max}=f^{\text{Kerr}}_{_{\text{ISCO}}}$  \cite{Favata:2010ic}, for ECOs we take the contact point.
The model signal and the true signal are parameterized by    $\theta^i=(\text{ln}~\mathcal{A},\text{ln} ~\mathcal{M},\text{ln} ~\eta, \Psi_c,t_c,\chi_1,\chi_2, \gamma_{_{ECO/UCO}})$, whose arguments are the amplitude $\mathcal{A}$, the chirp mass $\mathcal{M}=\eta^{3/5}M$, the symmetric mass ratio $\eta=M_1M_2/M^2$, the phase $\Psi_c$, the time at coalescence $t_c$, the dimensionless spin parameters  $\chi_1,\chi_2$ and the ECOs/UCOs dimensionless effective absorption coefficient given in Eqs.~(\ref{ge}),(\ref{guc}) respectively.

For this set of parameters, the root-mean-square error in measuring $\gamma_{_{ECO/UCO}}$ is expressed through the inverse of the Fisher matrix
\begin{gather}
	\sigma_{\gamma}=\sqrt{\langle \left(\Delta \gamma\right)^2\rangle}= \sqrt{\left(\Gamma^{-1}\right)_{\gamma \gamma}}~.
\end{gather}
For a  binary inspiral, the Fourier transform of the signal is modelled by  $\tilde{h}(f,\theta^{i})=\mathcal{A} e^{i \Psi}$, where
\begin{gather}
	\Psi=\Psi_{PP}+\gamma_{i}\Psi_{TH}~,
	\label{PS1}
\end{gather}  are the phases of the point-particle and tidal heating effects respectively \cite{Datta:2020gem}. $i=(\text{ECO,UCO})$ denotes the ECO/UCO effective absorption Eqs.~(\ref{ge}),(\ref{guc}). The phase is computed by solving the equation
\begin{gather}
	\dfrac{d^2 \Psi}{d f^2}~=~\dfrac{2 \pi}{\dot{E}} \dfrac{dE}{df}~,
\end{gather}
where $E=-M_1M_2/2b$ is the gravitational binding energy of the binary system.
The approximation method adopted here is the analytical ``TaylorF2 approximant'' \cite{Damour:2000zb,Arun:2004hn,Buonanno:2009zt}. We include correction terms to the GW phase in the form of spin-orbit, spin-spin and cubic spin corrections up to 3.5PN order relative to the leading-order GW term \cite{Isoyama:2017tbp,Khan:2015jqa}, and tidal heating correction term for spinning objects to the leading 2.5 PN order relative to the leading-order GW term \cite{Cardoso:2019rvt,Isoyama:2017tbp}. The amplitude is taken to leading PN order and includes the sky-averaged prefactor \cite{Berti:2004bd}.

\section{Discussion}

\subsection{Theoretical implications on black-hole mimickers}\label{tif}
In this section we discuss, on a qualitative level, the possible interpretations to a measurement of $\gamma_{_{UCO}}$ and its deviations in BH mimickers. Furthermore, we show explicitly how stringy inspired quantum corrections provides the deviation of the absorption coefficient from that of a classical BH.

 BHs-like objects are characterized by the BH compactness and an entropy that matches the Bekenstein's entropy $S_{_{BH}}=A/4$ \footnote{These objects can have Planckian corrections at the horizon scale, so $\epsilon=l_p$ and redshift arguments provides the Bekenstein's entropy $S_{_{ECO}}=1/\epsilon^2$ \cite{tHooft:1984kcu,Brustein:2018ixz}}.  This property is provided by Qunautm BHs (QBHs) models considered in \cite{Bekenstein:1995ju,Skenderis:2008qn,Almheiri:2012rt,Brustein:2016msz}.
We can express the UCO absorption Eq.~(\ref{guc}) in terms of their entropy (in units of $\hbar=k_B=1$), $\rho \nu_{_{BH}}=\eta_{_{BH}}=s_{_{BH}}/(4\pi)$, where $s_{_{BH}}=S_{_{BH}}/V$ is the entropy volume density. 
Then from Eq.~(\ref{guc}) at large viscosities $\gamma_{_{UCO}}=\eta_{_{UCO}}/\eta_{_{BH}}$, the absorption coefficient becomes
\begin{gather}
\gamma_{_{UCO}}~=~4 \pi \dfrac{\eta_{_{UCO}}}{s}~.
\end{gather}

So for UCO posses BH entropy, constraints on the absorption coefficient translates to constraints on the universal shear viscosity to entropy volume density, known as the KSS bound \cite{Kovtun:2004de} $4 \pi \frac{\eta_{_{UCO}}}{s}\geq1$. Then, since the UCO absorption is bounded from above ($\gamma_{_{UCO}}\leq 1$), the two conditions implies $\gamma_{_{UCO}}=1$.

The conjectured KSS bound found to agree with common substances as water, helium, and nitrogen. However, in strongly interacting systems such as quark-gluon plasma, which appear at extreme densities and are relevant for our understanding  of the UCO interior, the ratio found to be close to the bound and perhaps violating it \cite{Shuryak:2004cy}. Therefore, it is requested to explore the violation possibility ($\gamma_{_{UCO}}<1$) and its implications on the UCO absorption.

Let us assume that the fluid interior is composed of strongly interacting fluid in which, according to the fluid-gravity correspondence \cite{Rangamani:2009xk} their higher dimensional holographic 
dual is the stringy inspired $ADS_5$ black brane Gauss-Bonnet (GB) gravity. The Lagrangian density $\mathcal{L}=R-2\Lambda+\frac{\lambda_{{GB}}}{2}L^2\left(R^2-4R_{\mu\nu}R^{\mu\nu}+R_{\mu\nu\rho \sigma}R^{\mu\nu\rho \sigma}\right)$, where $L$ is the ADS radius and $\Lambda=-{6}/{L^2}$. $\lambda_{GB}$ is the dimensionless parameter that controls the magnitude of the higher curvature correction terms.
 In \cite{Brigante:2007nu,Kats:2007mq}, the bound was calculated from the holographic gravity dual at large $N$ and large 't Hooft coupling $\lambda$ ($1\ll \lambda\ll N  $), then to nonperturbative order in $\lambda_{{GB}}$ the result is given by
\begin{gather}
	\gamma_{_{UCO}}~=~1-4\lambda_{GB}
\end{gather}
Which can be expressed in terms of the reflection coefficient
\begin{gather}
		|\mathcal{R}|^2~=~4\lambda_{{GB}}
\end{gather}
This demonstrates how quantum corrections in the form of higher derivative terms modifies the BH absorption/reflection.
The positive sign of $\lambda_{GB}$ that is required in order to maintain causality \cite{Brustein:2017iet} is consistent with the positivity of $|\mathcal{R}|^2$ and the UCO absorption bound $\gamma_{_{UCO}}<1$.


\section{Summary}
In this paper, we investigated the effect of energy dissipation in ECOs at the inspiral phase. In particular, we were interested in whether and how can ECO mimic the effect of tidal heating in BHs. Then, in analogy to the membrane paradigm, by modelling ECOs as a Newtonian viscous body and UCOs as a viscous membrane, we find the corresponding absorption coefficients.
Then, the ECOs effective absorption was generalized to the spinning case,  according to Eq.~(\ref{ge}).

The detectability of the ECO effective absorption $\gamma_{_{ECO}}$ in future LISA observations was discussed. We performed a Fisher information analysis to estimate the statistical error in measurement of $\gamma_{_{ECO}}$ and to quantify how well can the ECO viscous interior produce similar effects to that of a BH. The results show that for supermassive, highly-spinning comparable-mass binaries, a relative absorption of $\gamma_{_{UCO}}\gtrsim 0.5,~\gamma_{_{ECO}}\gtrsim 0.9$ could be measured for binary at a luminosity distance $D_l=2$Gpc of $M=10^6 M_{\odot}$ and spin 
$\chi \gtrsim 0.75$ with more than $1 \sigma$ confidence, where at spins close to extremality $\chi\sim 0.95$, $\gamma_{_{UCO}}$ can be measured with more than $3\sigma$ confidence. 
Our results found to agree with \cite{Maselli:2017cmm}.
Nevertheless, a different approach for calculating the reflection coefficient was recently studied in \cite{Chen:2020htz}, which by 
quantifying the linear response of the membrane to the external tidal field, different boundary conditions are imposed on the membrane.



\section*{Acknowledgments}
I thank Ramy Brustein, Yoav Zigdon and Sayak Datta for discussions and comments on the manuscript and Kent Yagi for discussions.
The research was supported by the Israel Science Foundation grant no. 1294/16 and by the Negev scholarship.

\setcounter{equation}{0}
\renewcommand\theequation{A.\arabic{equation}}
\appendix
\section*{Appendix}
\section{Estimation of the neutron star shear viscosity}
\label{A1}

The estimated  density for  a typical neutron stars with $M~=~1.4\times M_{\odot}$, $R~=~10$km is
\begin{gather}
\rho~=~\dfrac{3}{4\pi}\dfrac{M_1}{R_1^3}~\approx~6.3\times 10^{17}\dfrac{kg}{m^3}\left(\dfrac{M}{1.4\times M_{\odot}} \right)\left(\dfrac{10^4m}{R} \right)^3~.
\end{gather}
The estimated viscosity for `cold' neutron star $T<10^9$K, where the viscosity is governed by the electron scattering, is given by \cite{LL1,LL2} 
\begin{gather}
\nu_{_{NS}} ~\approx~3\times 10 ^{5} \dfrac{m^2}{s}\left(\dfrac{\rho}{6.3\times 10^{17}\frac{kg}{m^3}} \right)\left(\dfrac{10^6 \text{K}}{T }\right)^2~.
\label{eta}
\end{gather}

	\section{The function $G(R,\omega,\nu)$}	
		For completeness we bring the full form of the function $G(R,\omega,\nu)$ Eq.~(\ref{bc}), for the case of our interest we consider zero bulk viscosity which found to be negligible when $\epsilon\ll1$. The function is taken from the Appendix of \cite{Maggio:2020jml} and defined by $G(R,\omega,\nu)=A/B$  

	\begin{equation}
\begin{split}
A~=~-&2 \epsilon  (49152 \pi ^2 \eta  w^2 (\epsilon -1)^5 (\eta  (4 \epsilon -5)-2)+\\&3 (-4 w^2 (4 \epsilon  (\epsilon  (4 \epsilon -9)+6)+5) (\epsilon -1)^2+\\&4 \epsilon  (2 \epsilon  (15-8 (\epsilon -1) \epsilon )-35)-7)-\\&512 i \pi  w (\epsilon -1)^3 \left(\eta  (\epsilon  (16 \epsilon -5)-29)+8 (\epsilon -1)^2\right))\\
B~=~&4 (7-4 \epsilon ) (\epsilon -1)^2 (8 w^2 (2 \epsilon  (4 \epsilon  (12 (\epsilon -4) \epsilon +71)-\\&181)+91) (\epsilon -1)^3+128 i \pi  \eta  w (8 \epsilon  (4 \epsilon  ((\epsilon -4) \epsilon +6)-\\&17)+11) (\epsilon -1)^3-8 \epsilon  (\epsilon  (2 \epsilon +3)-3)-11)~.
\end{split}
\end{equation}		
		Where $w=M\omega$ and $\omega=2\Omega$ is twice the orbital frequency. $\epsilon=\Delta R/R=(R-2M)/2M$ is the deviation from the BH horizon. In accordance with Eqs.~(\ref{bc}),(\ref{ruc}), since $G(\epsilon,w)\propto\epsilon$, in the BH limit $\epsilon\rightarrow0$ we get $\mathcal{R}=0,~ \gamma_{_{UCO}}=\gamma_{_{BH}}=1$ as expected.

	\end{subequations}
\end{document}